\documentstyle[aps,prb,epsfig]{revtex}
\def\narrowtext{} \tighten
\twocolumn

\begin{document}
\draft
\title{
Structure and Spin Dynamics of La$_{0.85}$Sr$_{0.15}$MnO$_3$}
\author{L. Vasiliu-Doloc$^{1,2}$, J. W. Lynn$^{1,2}$, A.H. Moudden$^{3}$, A.M. 
de Leon-Guevara$^4$ and A. Revcolevschi$^4$}
\address{$^{1}$NIST Center for Neutron Research, National Institute of 
Standards and Technology, Gaithersburg, Maryland 20899}
\address{$^2$Center for Superconductivity Research, University of Maryland, 
College Park, MD 20742}
\address{$^3$Laboratoire L\'eon Brillouin, CEA-CNRS, CE/Saclay, 91191 Gif sur 
Yvette Cedex, France}
\address{$^4$Laboratoire de Chimie des Solides, Universit\'e Paris-Sud, 91405
Orsay, France}
\address{%
\begin{minipage}[t]{6.0in}
\begin{abstract}
Neutron scattering has been used to study the structure and spin dynamics of 
La$_{0.85}$Sr$_{0.15}$MnO$_3$. The magnetic structure of this system is 
ferromagnetic below $T_C \simeq$ 235 K. We see anomalies in the Bragg peak 
intensities and new superlattice peaks consistent with the onset
of a spin-canted phase below $T_{CA} \sim 205$ K, which appears to be associated
with a gap at $q$ = (0, 0, 0.5) in the spin-wave spectrum.
Anomalies in the lattice parameters indicate a concomitant lattice distortion.
The long-wavelength magnetic excitations are found to be conventional spin 
waves, with a gapless ($<$ 0.02 meV) isotropic dispersion relation $E = Dq^2$. 
The spin stiffness constant $D$ has a $T^{5/2}$ dependence at low $T$, and the 
damping at small $q$ follows $q^4T^{2}$. 
An anomalously strong quasielastic component, however, develops at small 
wave vector above $\sim$ 200 K and dominates the fluctuation spectrum as 
$T \to T_C$. At larger $q$, on the other hand, the magnetic excitations become 
heavily damped at low temperatures, indicating that spin waves 
in this regime are not eigenstates of the system, while raising the temperature
dramatically increases the damping. The strength of the spin-wave damping also
depends strongly on the symmetry direction in the crystal. These anomalous 
damping effects are likely due to the itinerant character of the $e_g$ 
electrons. 
\typeout{polish abstract}
\end{abstract}
\end{minipage}}

\maketitle
\narrowtext

{\bf INTRODUCTION}\\

The correlated dynamics of spins and charges near the Mott transition in
doped lanthanum manganites La$_{1-x}$A$_{x}$MnO$_{3}$ \cite{jonker} has
generated continued interest, both because these systems exhibit anomalously
large magnetoresistance effects near the Curie temperature \cite{ecorr,cmr},
and because the physics of this class of materials is related to the 
high-T$_{c}$ superconducting copper oxides. Like the cuprates, these materials 
are in the vicinity of an insulator-metal transition, as well as magnetic and
structural instabilities, while to date no superconductivity has been found.
Rather, they exhibit exotic properties such as a dramatic increase in the
conductivity when the system orders ferromagnetically, either by cooling in
temperature or by applying a magnetic field. 
It has been suggested that an understanding of these materials must include,
in addition to the double exchange mechanism \cite{zener} and strong
electron correlations \cite{ecorr}, a strong electron-phonon interaction 
\cite{millis}. 
Cooperative Jahn-Teller (JT) distortions associated with the Mn$^{3+}$ JT ions
have been evidenced from structural studies at low doping, where the system is
insulating and antiferromagnetic, and may be an important contribution to
orbital ordering, double exchange, and related spin ordering and transport
properties observed at higher concentrations. As the doping concentration $x$
increases, the static JT distortion weakens progressively and the system becomes
metallic and ferromagnetic. 
It is believed that in the absence of a cooperative effect in this
regime, local JT distortions persist on short time and length scales. These
short-range correlations, together with the electron
correlations, would create the effective carrier mass necessary for large
magnetoresistance. 
The existence of strong electron correlations is expected to
affect the magnetic ordering and the magnetic excitation spectrum, and thus
the spin dynamics can provide crucial information for determining the
itinerancy of the system and the importance of the electron correlations.

A number of inelastic neutron scattering studies of the spin dynamics has been 
carried out in these systems, but most of them were carried out either in the
strongly-doped regime, or in the lightly-doped limit, and only a few of them
investigated the spin-wave dispersion to the zone boundary \cite
{martin,moudden1,jeff,perring}. These studies suggested essentially standard 
spin 
dynamics of a conventional metallic ferromagnet, except for the Ca-doped samples
\cite{jeff} and, more recently, for Nd$_{0.7}$Sr$_{0.3}$MnO$_3$ and 
Pr$_{0.7}$Sr$_{0.3}$MnO$_3$, \cite{baca} where coexistence of spin-wave 
excitations and a spin diffusion component was observed in the ferromagnetic 
phase. Perring et al. \cite{perring} used 
pulsed neutron scattering to measure the spin-wave dispersion throughout the
Brillouin zone of La$_{0.7}$Pb$_{0.3}$MnO$_{3}$, and concluded that a simple
isotropic Heisenberg Hamiltonian with only nearest-neighbor coupling could
account for the entire dispersion relation. They did not observe any
intrinsic linewidths at low temperature, possibly because of the coarse
resolution needed to observe the high energy spin waves. They did observe,
however, an unusual broadening of the high-energy spin waves at elevated
temperatures. 

At the other end of the ($x$,$T$) phase diagram, the system is semiconducting
for low doping concentrations,
with a canted spin structure \cite{moudden2,hirota}. The spin dynamics are 
highly anisotropic and are described by a strong in-plane ferromagnetic
exchange coupling and a weak out-of-plane antiferromagnetic coupling. The
ferromagnetic coupling in this limit results from orbital
ordering in agreement with Goodenough-Kanamori rules. Substituting La with Sr
introduces doped holes in the system, but also lattice distortions that
modify the overlap of the orbitals. Thus a clear competition arises between the
orbital-ordering inducing antiferromagnetic or ferromagnetic correlations and
the double exchange mechanism, which favors ferromagnetic correlations.
It is therefore of great interest to elucidate the crossover mechanism
from the weakly-doped to the strongly-doped limit. In the intermediate doping
regime we expect to see a strong deviation from the double-exchange model
prediction for the spin wave spectrum. 

So far, no results on the spin dynamics in the intermediate doping regime have 
been reported. We have therefore concentrated on 
La$_{0.85}$Sr$_{0.15}$MnO$_{3}$.
We find that the long-wavelength magnetic excitations are conventional spin 
waves at low $T$, while for $T \to T_C$ an anomalously strong quasielastic 
component develops at small wave vectors and dominates the fluctuation spectrum.
In the large wave vector regime very large intrinsic linewidths develop with 
increasing $q$ in the
ground state of this system, and we find that the strength of this damping
depends strongly on the symmetry direction in the crystal. This
demonstrates that conventional magnons are not eigenstates of this system at
large $q$, which is in contradiction to the expectations for a standard
Heisenberg model, or a simple one-band, fully-polarized double exchange
(half-metallic) ferromagnet. We also find that increasing the temperature
dramatically reduces the lifetime of the large-$q$ magnetic excitations,
indicating that there are strongly temperature-dependent contributions to
the spin-wave damping as well. These anomalous damping effects are likely
due to the itinerant character of the $e_{g}$ electrons. From the structural 
point of view, we 
observe anomalies in the Bragg peak intensities accompanied by the appearance of
superlattice peaks indexed with respect to the orthorhombic $Pbnm$ cell as 
(0 0 $l$) with $l$=odd or ($h \pm$ 1/2, $k \pm$ 1/2, 0) in related twin domains.
These are consistent with the onset of a spin-canted phase below $T_{CA}$ = 205 
K, as previously suggested by Kawano et al. \cite{kawano1}. The superlattice 
peaks are extremely weak, indicating that this subtle distortion
of the magnetic structure can be treated as a small perturbation of the purely 
ferromagnetic arrangement. Anomalies in the lattice parameters indicate a 
concomitant structural distortion. Other strong superlattice peaks, indexed
as ($h$, $k$, $l \pm 1/2$) or ($h \pm 1/4$, $k \pm 1/4$, 0) in related twin 
domains, have been reported by Yamada et al., and interpreted as the 
manifestation of a polaron lattice that would form below 190 K. We did not
observe such strong peaks in our $x$ = 0.15 sample using neutron scattering,
but we did observe weak and broad peaks using synchrotron X-ray scattering, 
which in addition presented strong irradiation effects. 
A comprehensive analysis of these superlattice peaks is in
progress \cite{wochner}.
We note that preliminary results of our studies have been reported previously
\cite{MMM96,APS}.
\\

{\bf EXPERIMENT}\\

The single crystal of La$_{0.85}$Sr$_{0.15}$MnO$_{3}$ used in the present
inelastic neutron scattering experiments was grown in Laboratoire de Chimie des 
Solides, Orsay, France, using the floating zone method. The crystal is 5-6 
mm in diameter and 3 cm in length and weighs 6.5 grams. It was
oriented such that the [010] and [001] axes of the orthorhombic $Pbnm$ cell
lie in the scattering plane. The inelastic measurements reported
here were performed on the BT-2, BT-4, BT-9 (thermal beam) and NG-5 (cold
source) triple-axis spectrometers at the NIST research reactor. In order to
reduce extinction effects, the diffraction experiments were carried out on a 
much smaller single crystal weighing 0.3 g, grown under the same conditions.
In all cases pyrolytic graphite
(PG) (0 0 2) has been used as an analyzer. As monochromator, the same
reflection was used on the BT-2, BT-9 and NG-5 spectrometers, while the copper
(2 2 0) reflection was used on BT-4 to investigate the higher energy 
excitations. Either PG or cold Be filters were used as appropriate, along with
a variety of collimator combinations. 
\\

{\bf STRUCTURE AND MAGNETIC ORDER}\\

La$_{0.85}$Sr$_{0.15}$MnO$_3$ has a high-temperature rhombohedral phase
($R\bar{3}c$), and undergoes a structural phase transition at $T_S$ = 360 K to 
an orthorhombic phase ($Pbnm$), as shown in Fig. 1(a).
In the paramagnetic phase the resistivity of 
this system initially increases with decreasing temperature \cite{uru,anane},
but then shows an abrupt drop associated with the onset of ferromagnetic
long-range order at $T_{C}$ = 235 K, as confirmed by magnetization measurements
versus $T$ obtained from Bragg scattering (see Fig. 1(b)). At a lower 
temperature around 205 K, the resistivity exhibits a fairly sharp upturn 
accompanied by anomalies in the Bragg peak intensities and the appearance of 
superlattice peaks, consistent with
a subtle distortion of the magnetic structure below $T_{CA}$ 
$\simeq $ 205 K \cite{kawano1} and a possible structural distortion 
\cite{wochner}, as described below. These superlattice peaks are weak, arising 
from a small perturbation of the FM structure, and
thus from the dynamical point of view this system is a ferromagnet to a good
approximation throughout the magnetically ordered state.

The integrated intensity of the (0 2 0) ferromagnetic Bragg reflection as a 
function of temperature is shown in Fig. 1(b) (closed squares). This
reflection has a weak nuclear structure factor, and therefore has a small 
intensity in the paramagnetic phase. Below $T_C$, magnetic scattering 
due to the ferromagnetism
of spins aligning on the manganese atoms produces a magnetic structure
factor and results in an increase of intensity. The data in Fig. 1(b)
are the same on warming and on cooling, with no visible hysteresis. It is thus
very likely that this ferromagnet exhibits a second-order phase transition.
Nevertheless, these data cannot be used to extract a critical magnetic exponent
$\beta$, because of the anomaly we see below 210 K. There is a loss of intensity
in the (0 2 0) reflection which coincides, on one hand, with the upturn 
in the resistivity and, on the other hand, with the appearance of weak 
superlattice peaks at antiferromagnetic Bragg points along the $c$ direction, 
such as the (0 0 3) reflection, also represented in Fig. 1(b) (closed 
circles). Note the factor of ten difference in intensity scales, with the 
(0 0 3)
intensity being more than an order of magnitude lower in strength. In the $Pbnm$
symmetry, (0 0 $l$) nuclear reflections with $l$=odd are not allowed, and we see
in Fig. 2 that there is no significant scattering present at Q = (0 0 3) at 215 
K. The appearance of such new reflections indicates that a second 
phase transition occurs at $T \sim$ 205 K, of magnetic and/or structural nature.
We can readily discard the possibility that the Bragg peaks at (0 0 $l$) with 
$l$=odd come from a second, distinct phase present in the sample, because the
205 K transition is also evident in the (0 2 0) reflection. 
These data are consistent with the onset of a canted spin structure at 205 K.
Indeed, the canting of spins would result in a small antiferromagnetic moment
at the expense of a reduction in the ferromagnetic moment (see Fig. 3). 
We estimate the cant angle to be $9.4^{\circ} \pm 0.8^{\circ}$.
In the double exchange model holes can only hop if adjacent spins are 
aligned, and the onset of ferromagnetism at 235 K gives rise to a metallic-like 
behavior of the resistivity. At 205 K, a residual superexchange interaction 
(left over from the low-doping limit) cants the spins
away from the perfect ferromagnetic arrangement. The AF component suppresses the
metallic state by reducing the matrix 
elements of electron hopping between Mn sites, thus explaining the upturn in the
resistivity. A similar behavior has been observed for $x$=0.1 and 0.125. 
\cite{kawano1}
The new type of magnetic long range order that occurs below $T_{CA}$ appears to 
be accompanied by the opening of a small gap at $\boldmath q$ = (0, 0, 0.5) in 
the spin-wave spectrum, and by a lattice distortion as described below.

Figure 1(a) shows the integrated intensity of the (1 2 0) nuclear Bragg 
reflection, which is allowed only in the orthorhombic symmetry. Therefore,
there is no intensity observable on cooling until
$T_S \sim 360$ K, where a sharp increase occurs when the system undergoes the 
rhombohedral-to-orthorhombic structural phase transition.
Upon cooling further, we see that the intensity of the (1 2 0) reflection starts
decreasing at $T_C$ = 235 K, reaches a local minimum at $T_{CA}$, and then 
starts increasing again. This  break in the intensity of the (1 2 0) reflection
corresponds closely to the anomaly in the ferromagnetic (0 2 0) peak and 
indicates a coupling between the magnetic system and the lattice.
Therefore, the new type of magnetic order that appears at $T_{CA}$ is 
accompanied by a lattice distortion, also evident in anomalies of the lattice 
parameters, as shown in Fig. 4(a). The $a$ and $b$ lattice parameters determined
from single-crystal diffraction are shown vs. $T$ in Fig. 4. Above $T_C$, the 
difference between $a$ and $b$ is large, indicating that the system is in the 
distorted-perovskite  orthorhombic O$^{\prime}$ phase. Upon approaching 
$T_C$, both $a$ and $b$ start decreasing abruptly, but have opposite variations
below $T_C$, with a maximum in $a$ and a minimum in $b$ half way between $T_C$ 
and $T_{CA}$. Below 200 K the difference between $a$ and $b$ becomes very small,
suggesting that the system releases the lattice distortions and goes into the 
undistorted-perovskite orthorhombic O$^*$ phase. Due to the uncertainty in 
assigning $a$ and $b$ from single-crystal diffraction on a twinned sample, it is
possible that the situation described in Fig. 4(b) is realized, where there is a
crossover between $a$ and $b$ at $T_C$. In this case, upon approaching
$T_C$, $a$ increases and $b$ decreases sharply, they cross at $T_C$, and 
continue their variation with the same slope until they reach a maximum and a 
minimum, respectively, before their variation changes sign. The upturn
in resistivity coincides with $T_{CA}$ and with the lower structural transition
temperature to the O$^*$ phase where $a$ and $b$ become equal, the same as 
observed for the $x$=0.125 system \cite{kawano2}. But unlike the
structural phase diagram proposed by Kawano et al. \cite{kawano2}, we 
find the $x$=0.15 system to be in the O$^{\prime}$ phase above $T_C$, not in the
O$^*$ phase. It is clear so far that the La$_{1-x}$Sr$_x$MnO$_3$ system in the 
concentration range $0.10 \leq x \leq 0.17$ is in the vicinity of structural
and magnetic instabilities. This only makes it more difficult to sort out all 
the subtleties of the magnetic and structural phase diagram in this region,
and further work may be needed to produce a reliable phase diagram for this 
system, for example by exploring a fine mesh of concentrations instead of 
drawing conclusions based on the study of a few selected doping levels.

The resistivity upturn for the $x$ = 0.10 and 0.15 systems has been associated 
by Yamada et al. \cite{yamada} with the onset of a polaron ordered phase with 
the periodicity $2a \times 2b \times 4c$ in the cubic perovskite cell. 
These authors have based their conclusion on the observation of superlattice 
peaks indexed as ($h \pm 1/4$,
$k \pm 1/4$, 0) or ($h$, $k$, $l \pm 1/2$) in a multidomain crystal composed of
microtwins. We have carried out a thorough search for such superlattice peaks 
using neutron scattering and have not found any. Therefore, we cannot confirm 
the model proposed by
Yamada et al. It is possible that this disagreement arises from slightly 
different nominal compositions and/or from the difficulty to index the numerous
reflections in a multidomain crystal, as the one used by Yamada et al. 
The possible existence of phases with compositional modulations in single
crystals grown using different techniques, or a short range order of the doping
element itself, could produce superlattice peaks as well. It is also known that 
these materials can easily incorporate vacancies which could lead to ordered 
defects. Using synchrotron 
X-ray scattering we observed weak and broad superlattice peaks at the positions 
indicated by Yamada et al., suggestive of a short-range order in the system, but
it necessitated the high flux from a synchrotron source to reveal them. 
The doping range $0.10 \leq x \leq 0.17$ is a critical concentration regime in 
the sense that the magnetic and nuclear structure vary rapidly with $x$. 
Nevertheless, the average nuclear structure of this system has been well 
established. This is why it is surprising that Yamada et al. 
present data showing (1 0 0) reflections in the orthorhombic phase of their 
samples, much stronger than the (2 0 0) reflection. In the orthorhombic $Pbnm$ 
symmetry this type of reflection is not allowed.

In conclusion, we believe that the upturn in the resistivity corresponds to the
onset of an AF 
component at $T_{CA}$. We emphasize that the AF component is small,
arising from a small perturbation of the FM structure, and thus the spin
dynamics of the $x$=0.15 system of present interest can be treated to a good 
approximation as a ferromagnet throughout the magnetically ordered phase.
\\

{\bf SPIN DYNAMICS}\\

a) {\bf Small wave vector regime} \\

We have investigated the spin dynamics in the (0,1,0) and (0,0,1)
high-symmetry directions of the orthorhombic unit cell. The ground state spin 
dynamics for a half-metallic
ferromagnet was not expected to differ much from the conventional picture of
well defined spin waves, and we found that the long wavelength magnetic
excitations were in fact the usual hydrodynamic spin waves, with a
dispersion relation given by $E=E_{0}+Dq^{2}$, where $E_{0}$ is the spin-wave
energy gap and $D$ is the spin stiffness constant. 
In order to set an upper limit on the
value of any energy gap $E_0$, we have performed very high-resolution inelastic
measurements on the NG-5 (SPINS) cold-neutron triple-axis spectrometer.
Figure 5(a) illustrates the 
measured spin-wave dispersion relation along the (0,0,1) direction at 10 K and 
220 K. The spin-wave energies have been obtained from least-squares fits of the 
data to the dispersion
relation convoluted with the instrumental resolution. The solid
curves are fits to $E=E_{0}+Dq^{2}$. The fitted value of $E_{0}$ at 10 K is
0.019 $\pm $ 0.04 meV, where the quoted error includes only statistical
errors. The negligible value of $E_0$ indicates that this material is a 
prototypical isotropic 
ferromagnet, comparable to the soft amorphous ferromagnets. Previous studies 
have shown that the undoped system is a Jahn-Teller 
distorted perovskite that consists of sheets of ferromagnetic spins in the $ab$
plane that are coupled antiferromagnetically along the $c$-axis. The spin-wave
energy gap is $\sim$ 2.5 meV for $x$=0, and then decreases continuously with $x$
and becomes zero for $x \simeq $ 0.1 as the antiferromagnetic $c$-axis
interaction changes sign and becomes ferromagnetic \cite{moudden2,hirota}.
The present measurements confirm that $x$=0.15 belongs to the ferromagnetic 
doping concentration range for which the spin-wave energy gap is essentially 
zero. The fitted values for the spin stiffness constant $D$ are 94.87 
$\pm$ 1.18 meV $\AA^2$ at 10 K, and 40.53 $\pm$ 0.13 meV $\AA^2$ at 220 K.
The low-temperature value of the spin stiffness constant gives a ratio of 
$D/k_{B}T_{c}$ $\sim $ 4.6 $\AA ^{2}$.
We can also see that energies for $q>$ 0.15 depart from the $q^{2}$-dependence,
as higher-order terms in the power law expansion of the dispersion relation
become significant, as expected.

Similar data have been obtained along the (0,1,0) direction, and
Fig. 5(b) plots the temperature dependence of the stiffness $D$ in both 
directions. This result shows that the exchange interactions (as estimated from
the long-wavelength spin dynamics) are
remarkably isotropic, and demonstrates again that the
dimensional crossover from the 2D-like planar ferromagnetic (in the A-type
AF ordering) to the 3D isotropic ferromagnetic character takes place at
concentrations lower than $x$=0.15, probably $x \sim $ 0.1, 
\cite{moudden2,hirota} where the system still behaves like a semiconductor. 
It is thus interesting to note that the metallic ferromagnetic features in the
spin dynamics appear at lower doping concentrations than the compositional
insulator-metal 
transition at $x \sim$ 0.17, above which the canting transition 
is no longer observed at any $T$.
This observation supports the conjecture that the 
insulator-metal phase boundary is actually located at $x$=0.1 in the
La$_{1-x}$Sr$_x$MnO$_3$ system, but the AF component suppresses the metallic 
state in samples with $0.1 \leq x \leq 0.15$, without having a strong
influence on the character of the spin dynamics.

The long-wavelength spin wave data can be compared to the Dyson formalism of 
two-spin-wave interactions in
a Heisenberg ferromagnet \cite{dyson}, which
predicts that the dynamical interaction between the spin waves gives, to
leading order, a $T^{5/2}$ behavior: 
\begin{equation}
D(T)=D_0\left\{ 1-\frac{v_{0}\overline{l^{2}}\pi }{S}\left( \frac{k_{B}T}{4\pi
D_0}\right) ^{5/2}\zeta (\frac{5}{2}) + \cdots\right\} ,  
\label{T5/2}
\end{equation}
where $v_{0}$ is the volume of the unit cell determined by nearest
neighbors, $S$ is the average value of the manganese spin, and 
$\zeta (\frac{5}{2})$ is the Riemann zeta integral. $\overline{l^{2}}$ is the 
moment defined by 
$\overline{l^{n}}=\frac{S}{3D}\left\{ \sum l^{n+2}J({\bf l})\right\} $ 
which gives information about the range of the exchange interaction. The solid 
curves in Fig. 5(b) are fits to Eq.~\ref{T5/2}, and are in good agreement with 
the experimental data for temperatures up to 200 K, surprisingly close to $T_C$.
The fitted values of 
$\overline{l^{2}}$ give $\sqrt{\overline{l^{2}}_{010}}$ = (2.37 $\pm
0.39)b_{0}$, and $\sqrt{\overline{l^{2}}_{001}}$ = (2.56 $\pm 0.75)c_{0}$,
where $b_{0}$ and $c_{0}$ are here the distances to the nearest neighbors 
($ \simeq $ 3.896 $\AA $), and indicate that the exchange interaction extends
significantly beyond nearest neighbors in both directions. For $T >$ 200 K the
experimentally measured values of $D$ depart from the leading order $T^{5/2}$ 
dependence as expected, having rather a power law behavior (dashed curves) and 
appearing to collapse as $T \to  T_{C}$.

In the course of these measurements we have noticed that 
the central quasielastic peak has a strong temperature dependence, while 
typically the central peak originates from weak temperature-independent nuclear 
incoherent scattering. Figure 6(a) shows two
magnetic inelastic spectra collected at 210 and 220 K, and
reduced wave vector $q$ = 0.125 away from the (0 0 2) reciprocal point. A flat
background of 0.6 counts plus an elastic incoherent nuclear peak of 100 counts,
measured at 10 K, have been subtracted from these data. 
We can clearly see the development of the quasielastic component, comparable in
intensity to the spin waves, and the strength of this scattering is shown in 
Fig. 6(b) as a function of temperature. We observe a significant intensity
starting at $\sim$ 210 K ($\sim$ 25 K below $T_C$), and the scattering peaks at 
$T_C$. At and above $T_C$ all the scattering is quasielastic. 
The appearance in the ferromagnetic
phase of a quasielastic component comparable in intensity to the spin waves
was first observed on Ca-doped polycrystalline samples \cite{jeff} and it has 
been suggested that it is associated with the localization of the $e_g$ 
electrons on the Mn$^{3+}$/Mn$^{4+}$ lattice, and that it may be related to the
formation of spin polarons in the system \cite{jeff2}. We have observed a 
similar anomalous behavior of the central peak in the strongly-doped system
La$_{0.70}$Sr$_{0.30}$MnO$_3$, \cite{sr30} but not in the pyrochlore 
Tl$_2$Mn$_2$O$_7$. \cite{pyrochlore}
It thus appears that the coexistence of spin-wave excitations and spin
diffusion is a common characteristic for many perovskite manganites, and it 
raises the question about its relevance for the colossal magnetoresistance 
property of the perovskite manganites. \\

b) {\bf Large wave vector regime} \\

One of the major open questions in the field of doped manganites concerns
the appropriate spin Hamiltonian in the ordered state. The
simplest candidate is the Heisenberg form with couplings $J_{ij}$ between
pairs of localized spins at sites {\bf R}$_{i}$ and {\bf R}$_{j}$, 
$H=-\sum_{ij}J_{ij} {\bf S}_{i}\cdot {\bf S}_{j}$. The double exchange model
in the limit of large Hund's coupling \cite{furu1} gives the same (cosine-band)
dispersion relation as a ferromagnetic Heisenberg model with nearest-neighbor 
spin exchanges, but is expressed in terms of different parameters (electron 
transfer energy $t$ and Hund's coupling $J_{H}$, $J_{H}/t \to \infty$ in this 
case). In the case of $J_H$ finite,
however, the double exchange model does not provide an analytical form of the 
dispersion relation that can be readily compared to the experimental data. For
simplicity reasons, we have compared our data to the Heisenberg model.

The spin-wave excitations have been measured to the zone boundary, and the 
dispersion curves at 10 K for the (0,1,0) and (0,0,1) directions are plotted in 
Fig. 7. The solid curves are the Heisenberg model with a ferromagnetic
nearest-neighbor coupling $J_{1}S$ = 3.55 $\pm $ 0.06 meV obtained from a fit 
in which all the data were given the same weight. If we perform a fit giving 
more weight to the low-$q$ data (where the spin-wave excitations are well 
defined and the energies have been obtained from high-resolution measurements), 
we obtain $J_{1}S$ = 3.00 $\pm $ 0.02 meV, and the result is depicted by the 
dashed curves. The magnon bandwidth in a double exchange model for large Hund's
coupling \cite{furu1} $J_H \to \infty$ would then be $E_{SW} \equiv 24J_{1}S$ = 
85.2 meV. Using a spherical free-electron model Fermi surface, we estimate an
electron transfer energy $t$ = 0.11 eV.
The overall agreement is reasonable in the first case (solid curves), except 
for a region of intermediate energies below $q$=0.5, while
in the second case (dashed curves) the agreement is excellent at low energies,
but becomes poor for $q \geq 0.5$.
In the (0,0,1) direction we see the manifestation of an additional feature:
the opening of a small gap in the spin-wave spectrum at $q=0.5$, which is
likely related to the presence of the odd-integer superlattice peaks at 
(0 0 $l$) that appear below $T_{CA}$ = 205 K. It would be interesting to check 
if this gap goes away above $T_{CA}$, when the system recovers the exact FM 
structure. Unfortunately, measuring spin waves above 100 K
proves to be difficult in this wave vector regime because, as we discuss 
below, raising the temperature dramatically increases the spin-wave damping to
the point that the larger-$q$ magnons become ill defined for temperatures in 
the range of $T_{CA} \sim 205$ K.

The overall agreement between the calculated and observed dispersion relation
might suggest that the Heisenberg model, or a single-band double exchange 
model, provides a good description of the spin dynamics of the manganite 
systems \cite{perring}. However, in both models the magnons in the
ground state are eigenstates of the system. The {\it observed} magnetic
excitations, interestingly enough, develop large intrinsic linewidths with
increasing $q$, even at low temperatures. Figures 8(a), (b) and (c) show for
comparison energy scans at different $q$-vectors along (0,1,0) measured at 10 
K, well below T$_{C}$ (235 K). The instrumental energy resolution (taking into 
account the slope of the measured dispersion relation) is 1.1 meV, 3.1 meV, 5.6
meV and 5.9 meV at $q$ = 0.2, 0.3, 0.6 and 1.0, respectively, while the 
observed widths are 2.4, 4.9, 28.8, and 46.2 meV, respectively.
To obtain quantitative linewidths, all the data have been fit to a 
convolution of the instrumental resolution function with the spin-wave cross 
section. For the very large linewidths found at large $q$, it is 
important to correct the data for the instrumental background, which is the sum
of the sample background, the fast
neutron background and the angle-dependent background for small scattering
angles (which represented a real problem only for scattering angles $< 
15^{\circ}$). The fast neutron background (dependent on the counting time which
can be different at different energy transfers) and the angle-dependent 
background have been determined by misorienting the analyzer crystal by 
5$^{\circ}$, and then this background has been subtracted from the data. 
The $Q$-independent sample background has been kept fixed in the
fits. At each $Q$, we have scanned a wide energy range to determine the
background. The intrinsic linewidths obtained from the fits, assuming a
damped harmonic oscillator spectral weight function \cite{lor},
are shown in Fig. 9(a). At the zone boundary in the (0,1,0) direction the 
damping of the magnetic excitation is $\sim 47$ meV, comparable
to the energy of the excitation, while the resolution width is 5.9 meV. 
Note that there are significant
linewidths at relatively small $q$ as well, but these are dwarfed by the huge
widths observed as one proceeds towards the zone boundary. The most likely
explanation for the observed damping is a strong magnon-electron
interaction due to the itinerant nature of the $e_{g}$ electrons associated
with the electron hopping and double-exchange mechanism. However, in a
simple single-band double exchange model the ground state is fully
polarized, and the magnon-electron interaction is forbidden by symmetry. 
This means that the single-particle Stoner-like excitations have a gap due to 
the energy difference between the up-electron and down-electron bands (case of 
a strong ferromagnet). Therefore, the 
continuum of Stoner modes is in a higher energy
region of the spin excitation spectrum, and the low-lying spin wave excitations
should be well-defined. Only at finite temperature does a gapless mode of 
Stoner
excitations appear, and this is expected to have a strong damping effect on the
spin wave excitations near the zone boundary \cite{furu1}. The
presence of the small superlattice peaks might also have an effect on the 
magnon linewidths by breaking the half-metallic symmetry, but it seems very
unlikely that this could provide an explanation for these extraordinarily
large linewidths at large $q$. Another possible contribution to the
spin-wave damping could come from an alloy effect related to the random
distribution of Mn$^{3+}$ and Mn$^{4+}$ ions and consequent distribution of
exchange interactions, but typically the linewidths due to alloying are
quite modest and not strongly anisotropic. Macroscopic inhomogeneities of the 
sample are ruled out since the magnetic and structural transitions are observed
to be sharp. Measurements in the (0,0,1) direction, on the other hand, reveal 
only modest linewidths for these spin waves, as also shown in Fig. 9(a), and 
similar to those observed very recently 
by Hwang et al. in Pr$_{0.63}$Sr$_{0.37}$MnO$_{3}$. \cite{Hwang} The strong 
damping and its directionality that we observe suggests that 
hybridization effects in a multiband model must be included in the description 
of these materials to quantitatively explain the spin dynamics, similar to 
conventional itinerant electron systems \cite{cooke}. 

As we mentioned before, there are significant linewidths at relatively small 
$q$ as well. In the long wavelength regime the linewidths in spin wave 
theory, due to spin wave-spin wave interactions, are expected to follow 
$\Gamma (q,T)\propto q^{4}\left[ T\ ln\left( kT/E_{sw}\right) \right]^{2}$. 
Figure 9(b) shows the experimentally observed spin-wave linewidths for 
La$_{0.85}$Sr$_{0.15}$MnO$_{3}$ for $q \leq 0.25$ $\AA^{-1}$ and for 
temperatures between 10 and 230 K. We see that the points (solid circles) for 
$q \leq 0.16$ $\AA^{-1}$ fall quite close to a single universal curve.
This indicates that spin wave interactions dominate 
the linewidths in this regime. At higher $q$, represented by the open squares in
this figure, the data deviate strongly from the expected straight line,
and are clearly more strongly damped, even at low temperatures. This is 
an interesting result in the following sense: The 
$q^{4}\left[ T\ ln\left( kT/E_{sw}\right) \right]^{2}$ dependence predicted by 
the linear spin wave theory is the leading-order behavior expected only in the 
small-$q$ regime. For higher $q$, this relation overestimates the observed 
linewidths, {\it i.e.} observed linewidths
plotted vs. $q^{4}\left[ T\ ln\left( kT/E_{sw}\right) \right]^{2}$ are expected
to lie underneath the straight line.  Our results show that the linear spin wave
theory relation holds up to $q \simeq 0.16$ $\AA^{-1}$, while for higher 
$q$ values the linewidths are found to sit {\it above} the expected straight 
line, not below. We remark that for larger $q$ ($> 0.26$ $\AA^{-1}$) the 
linewidths are too large to be included on the plot.

We finally note that increasing the temperature also has a strong effect on the
spin-wave damping. 
At $T$ = 50 K, still well below $T_{C}$, the zone-boundary magnetic 
excitation appears to be the same as at 10 K, but at 100 K it broadens to the 
point of being ill defined (with this
resolution), as can be seen in Fig. 8(d). This behavior contrasts markedly
with the expectations for a conventional localized-spin ferromagnet, and
indicates that there are additional strongly temperature-dependent
contributions to the spin-wave damping. The leading-order spin wave damping
in the double-exchange model is proportional to ($1-m^{2}$), where $m$ is
the reduced magnetization \cite{furu2}, and at this low a temperature the
expected temperature dependence again appears to be too small to explain
these observations. A quantitative description of the spin wave damping 
as a function of wave vector and temperature in these materials represents a 
theoretical challenge. \\

{\bf DISCUSSION}\\

A mean-field calculation gives for the Curie temperature of a localized 
ferromagnet a value $4J_{1}S(S + 1)$. If we use for $S$ the mean spin on the 
manganese ions, $S = \frac{3}{2} x + 2(1-x)$, we calculate a mean field value 
$T_C^{MF}$ = 482 K.
This is more than double the experimentally measured value. But it is
well known that fluctuations reduce the Curie temperature even for 
three-dimensional local moment ferromagnets. In particular, for the simple cubic
nearest-neighbor Heisenberg ferromagnet, fluctuations reduce $T_C$ to
$J_{1}[2.90S(S + 1) - 0.36]$. \cite{rush} We evaluate this quantity to be 341 K,
still 45\% larger than the measured value of 235 K.
In other words, the Curie temperature is inconsistent with the magnon bandwidth 
$E_{SW} \equiv 24J_{1}S$. For La$_{0.85}$Sr$_{0.15}$MnO$_3$ we find that the 
magnon bandwidth is significantly (about a factor of two) larger 
than the Curie temperature of 235 K. The large value of $J_1$ determined from 
our neutron scattering measurements of the spin wave spectrum indicates that 
this system is not localized, but has itinerant character.
The renormalization of $T_C$ from the mean-field value in a Heisenberg 
Hamiltonian (which is appropriate for ferromagnetic insulators or localized 
spin systems) is considered to be a good measure of the itineracy. An itinerant
ferromagnet will have a lower $T_C$ compared to the mean-field value, but will
have large values of $J$, consistent with the present results.
A calculation of $T_C$ performed 
by Furukawa within the double exchange model (the Kondo lattice model with 
ferromagnetic couplings) in the infinite-dimensional approach \cite{furu3} 
reproduces the experimental result for La$_{0.85}$Sr$_{0.15}$MnO$_3$
when using a bandwidth of the itinerant $e_g$ electrons $W$ = 1.05 eV and a 
Hund's coupling $J_H$ = 4.2 eV. This value of $W$ gives a large value of the 
electron transfer energy $t \equiv W/6 = 0.175$ eV, which is probably a more 
accurate value than the one we have previously estimated from the magnon
bandwidth using a spherical free-electron model Fermi surface.

This manifestation of the itinerant character of the spin system and  
the isotropy of the exchange interactions 
demonstrate that the dimensional crossover from the 2D-like
planar ferromagnetic in the low doping limit to the
3D isotropic ferromagnetic character takes place at concentrations lower than
$x$=0.15, probably $x \sim 0.1$. \cite{moudden2,hirota} It is remarkable
that the metallic ferromagnetic features in the spin dynamics appear at 
lower doping concentrations than the compositional insulator-metal 
transition at $x \sim$ 0.17. It is therefore conjectured that the 
insulator-metal phase boundary would have been located at $x$=0.1 in the 
La$_{1-x}$Sr$_x$MnO$_3$ system, if it were not for the spin canting transition 
that suppresses the metallic conduction in samples with $0.1 \leq x \leq 0.15$.
\cite{kawano1} It is this itinerant character of the $e_g$ electrons, viewed as
a hopping conduction channel implemented by the scattering of a high-energy 
spin wave, that may be responsible for the anomalous spin-wave damping effects 
observed in this system.
\\

Research at the University of Maryland is supported by the NSF under Grant
DMR 97-01339 and by the NSF-MRSEC, DMR 96-32521. Experiments on the NG-5
spectrometer at the NIST Research Reactor are supported by the NSF under
Agreement No. DMR 94-23101.

\newpage
\clearpage

{\large FIGURE CAPTIONS}\\

FIG. 1 (a) Temperature dependence of the (1 2 0)
nuclear Bragg reflection which is allowed only in the orthorhombic structure. 
The system undergoes an abrupt R-to-O$^{\prime}$ structural phase transition at 
$T_S \simeq 360$ K. The (1 2 0) reflection reacts both at $T_C$ and $T_{CA}$, 
indicating a coupling between the lattice and the magnetic system. (b)
Temperature dependence of the (0 2 0) ferromagnetic peak and (0 0 3) 
AF peak. Note that the onset of AF intensity occurs at $T_{CA} < T_C$ and 
coincides with a break in the ferromagnetic intensity consistent with a 
reduction in the FM moment due to a canting of spins. Note also the factor of 
ten difference in intensity scales, with the (0 0 3) intensity being about 40 
times weaker than the (0 2 0) intensity. 

FIG. 2 Temperature dependence of the profile of the (0 0 3) reflection. There is
no significant scattering at 215 K ($< T_C$ = 235 K).

FIG. 3 Schematic illustration of the magnetic structure of 
La$_{0.85}$Sr$_{0.15}$MnO$_3$: (a) purely ferromagnetic for $T_{CA} \leq T \leq
T_C$; (b) spin canted below T$_{CA}$. The cant angle is small, 
($9.4^{\circ} \pm 0.8^{\circ}$), resulting in a small antiferromagnetic 
component.

FIG. 4 Temperature dependence of the $a$ and $b$ lattice parameters showing the 
strong reaction of the lattice at $T_C$ and $T_{CA}$. $a$ and $b$ are very 
different above $T_C$ (the system is in the orthorhombic O$^{\prime}$ phase), 
then the ferromagnetism drastically suppresses the static lattice distortion and
$a$ and $b$ become almost equal at $T_C$, but next the precursor effects of the
phase transition at $T_{CA}$ distort the lattice once more, before releasing
the lattice distortion almost completely below $T_{CA}$. Due to the uncertainty 
in identifying $a$ and $b$ from single-crystal diffraction on a twinned sample,
the situation described in (b) may be realized, where there is a crossover 
between $a$ and $b$ at $T_C$.

FIG. 5 (a) Spin-wave dispersion along (0,0,1) in the long wavelength limit
at 10 K (open circles) and 220 K (closed circles). Solid curves are fits to 
$E=E_{0}+Dq^{2}$. (b) The spin-wave stiffness coefficient in the (0,1,0) (closed
circles) and (0,0,1) (open circles) directions vs. $T$. The solid curves are 
fits to Eq. (1). For $T >$ 200 K the measured values of $D$ depart from the 
$T^{5/2}$ dependence and the dashed curves are fits to a power law. 

FIG. 6 (a) Magnetic inelastic spectra collected at 210 and 220 K, and a
reduced wave vector $q$ = (0, 0, 0.125). A flat background of
0.6 counts plus an elastic incoherent nuclear peak of 100 counts, measured at
10 K, have been subtracted from these data. The dominant effect is the
development of a strong quasielastic component in the spectrum. (b) Integrated
intensity versus temperature for the central peak at $q$ = (0, 0, 0.15). The
quasielastic scattering starts increasing in intensity well below $T_C$.
Above $T_C$, all the scattering in this range of $q$ is quasielastic.

FIG. 7. Spin-wave dispersion along (0,1,0) and (0,0,1) measured to the zone 
boundary at 10 K. The solid curves are fits to a Heisenberg model with 
$J_{1}S = 3.55 \pm 0.06$ meV obtained by giving all the data the same weight.
The dashed curves are fits to a Heisenberg model with $J_{1}S = 3.00 \pm 0.02$ 
meV obtained by favoring the low-$q$ data.

FIG. 8. Magnetic inelastic spectra collected at 10 K for reduced wave
vectors (a) $q$ = 0.2 (open circles) and $q$ = 0.3 (closed circles), (b) $q$
= 0.6, (c) $q$ = 1.0 in the (0,1,0) direction. The magnetic excitations are 
increasingly broader with $q$. The solid curves are fits to a convolution of the
instrumental resolution function with the spin-wave cross section. The 
horizontal arrows indicate the width of the instrumental energy resolution 
E$_R$. (d) Energy scan at $q$ = 1.0 at 100 K, still well
below $T_{C}$, showing that the magnetic excitation at the zone boundary has 
broadened beyond recognition (with this resolution).

FIG. 9 (a) Intrinsic spin-wave linewidths versus $q$ at 10 K along the (0,1,0) 
(closed circles) and along the (0,0,1) direction (open squares). The damping 
at the zone boundary in the (0,1,0) direction is $\sim 47$ meV, comparable to 
the spin-wave energy, but it is smaller in the (0,0,1) direction. (b)
Intrinsic spin-wave linewidths versus the expected ($q,T$) dependence in 
linear spin wave theory. Linewidths of spin waves for $q\leq $ 0.16 $\AA ^{-1}$
(closed circles) fall quite close to a single universal curve, whereas
linewidths for $0.16<q<0.25$ $\AA ^{-1}$ (open squares) deviate
significantly from the expected straight line. The low temperature
linewidths at higher $q$ are off scale on this plot.

\newpage
\clearpage

\widetext

\vspace*{-1.5cm}
\epsfig{file=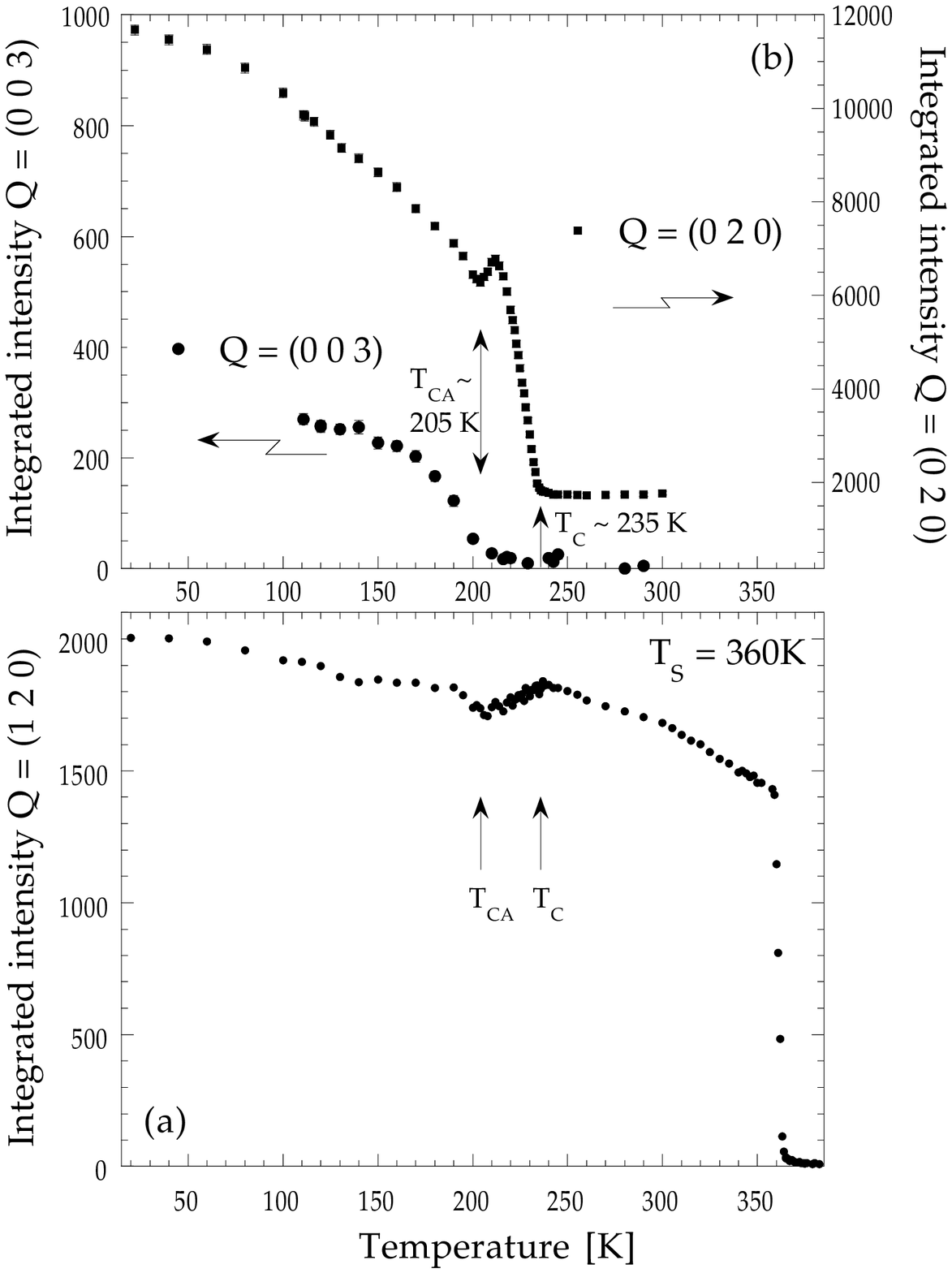,width=17truecm}

\Large
\hspace{8cm} FIG. 1: L. Vasiliu-Doloc et al.

\newpage
\clearpage
\vspace*{3.0cm}
\hspace{1cm} \epsfig{file=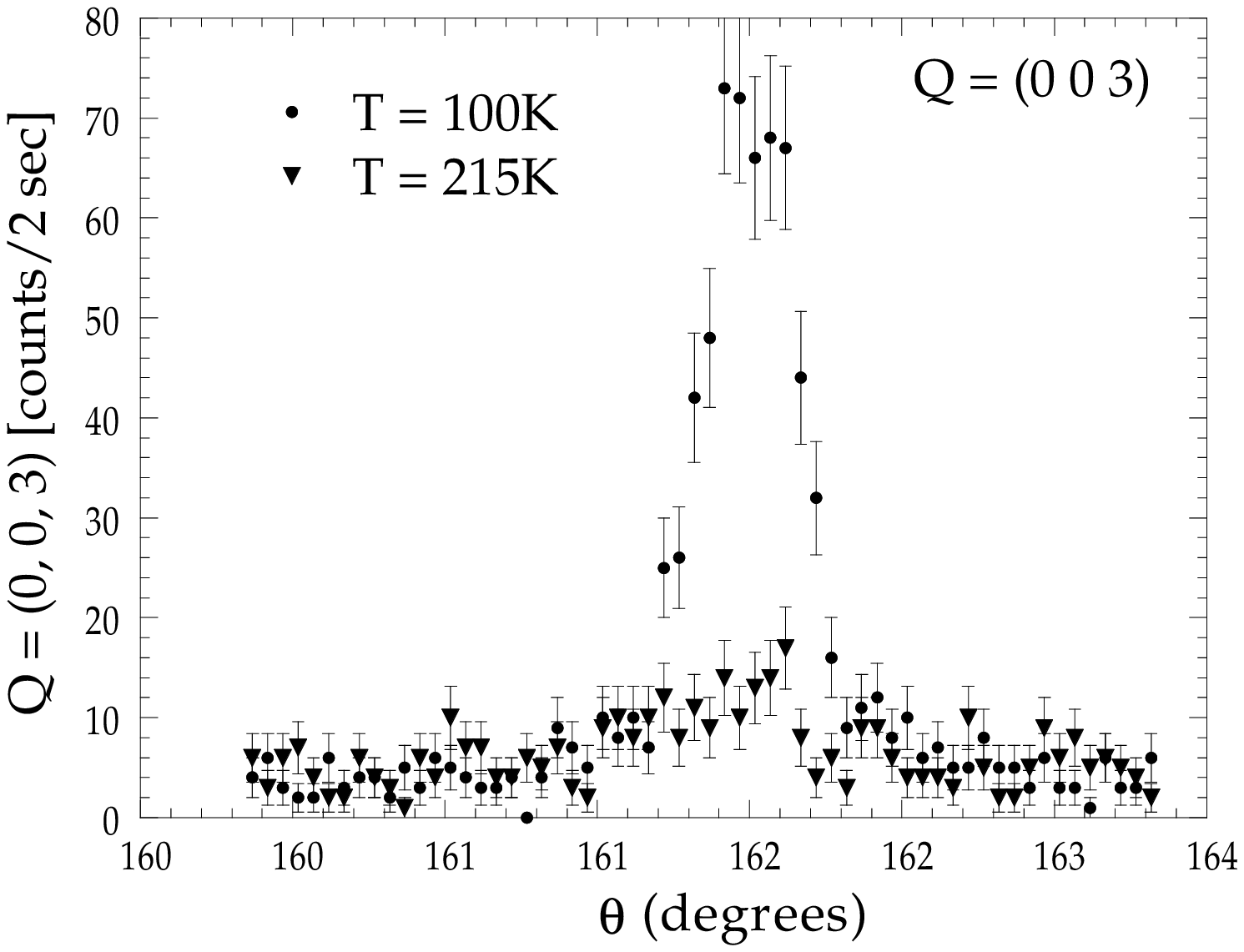}

\Large
\vspace*{2.0cm}
\hspace{8cm} FIG. 2: L. Vasiliu-Doloc et al.

\newpage
\clearpage
\vspace*{3.0cm}
\hspace*{1cm} \epsfig{file=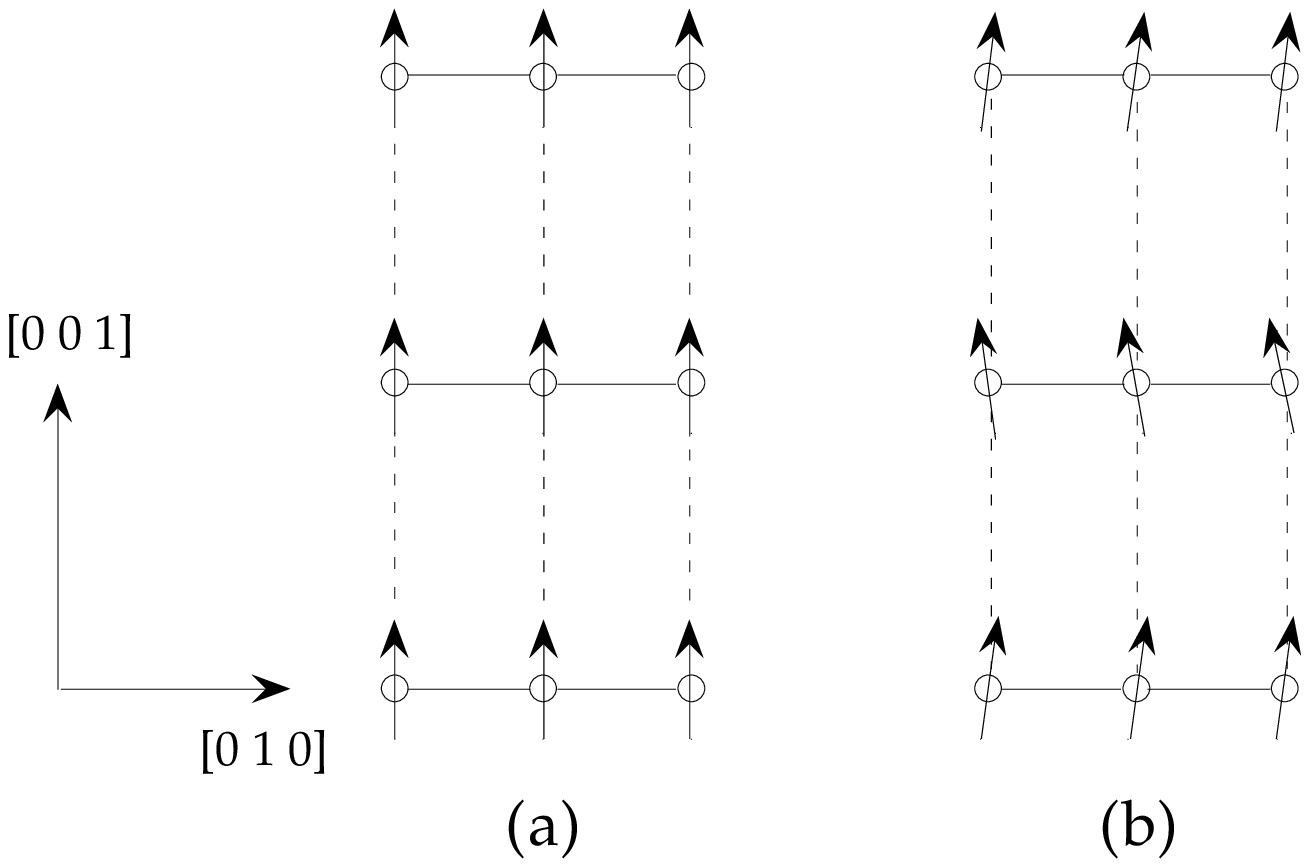}

\Large
\vspace*{2.0cm}
\hspace{8cm} FIG. 3: L. Vasiliu-Doloc et al.

\newpage
\clearpage
\vspace*{-1.5cm}
\hspace*{1cm} \epsfig{file=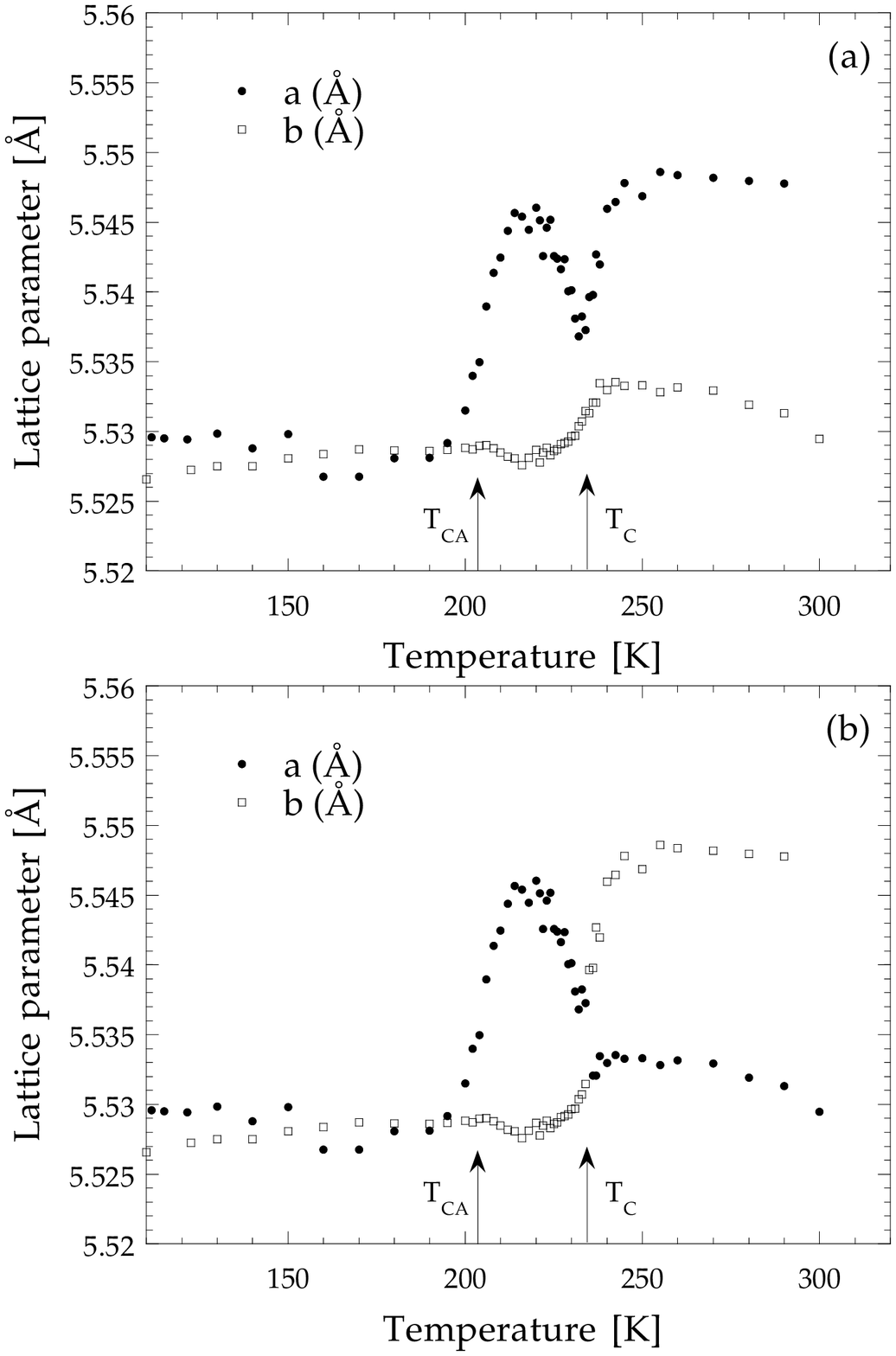,width=17truecm}

\Large

\hspace{8cm} FIG. 4: L. Vasiliu-Doloc et al.

\newpage
\clearpage
\vspace*{-1.5cm}
\hspace*{1cm} \epsfig{file=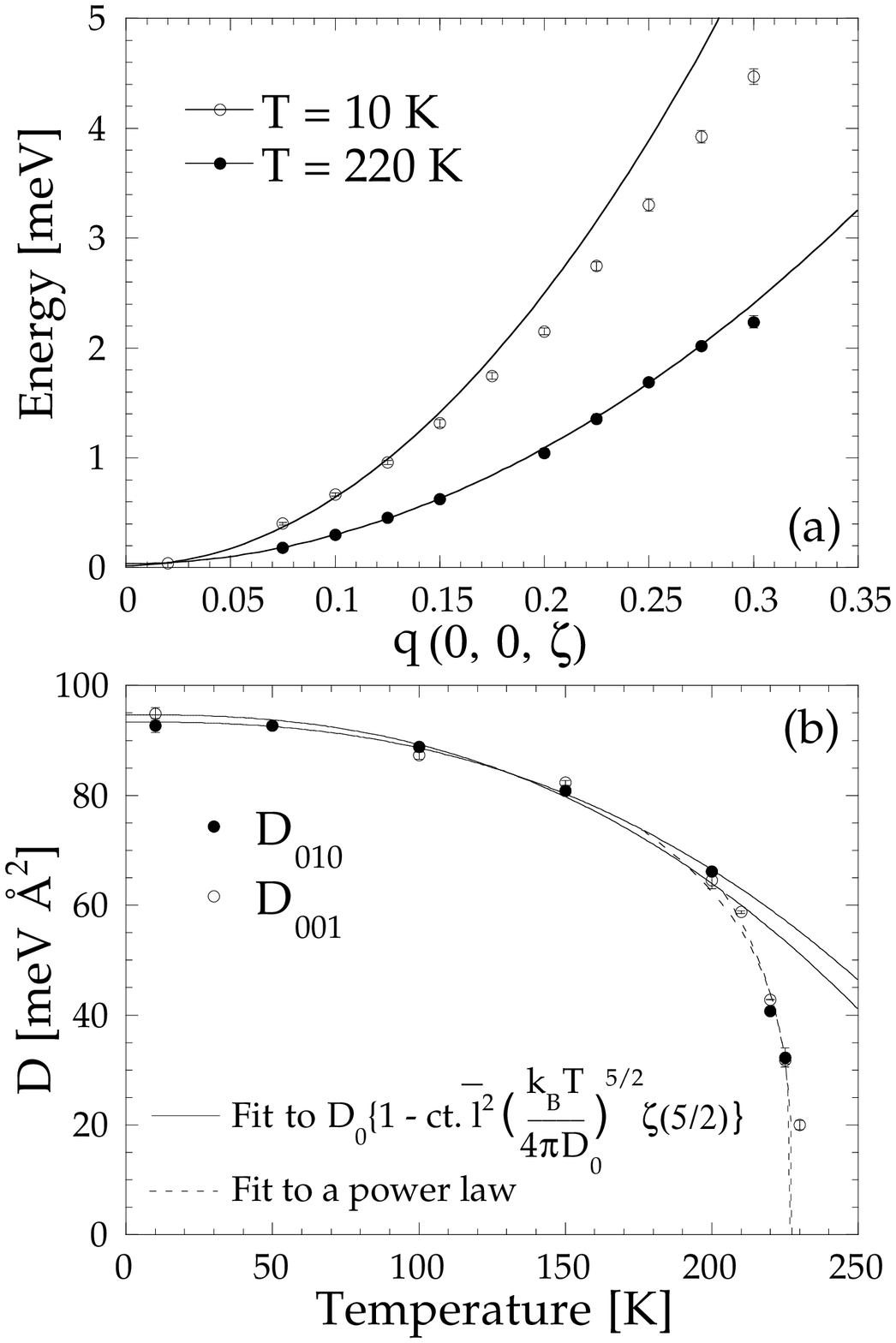,width=17truecm}

\Large
\hspace{8cm} FIG. 5: L. Vasiliu-Doloc et al.

\newpage
\clearpage
\vspace*{-1.5cm}
\hspace*{1cm} \epsfig{file=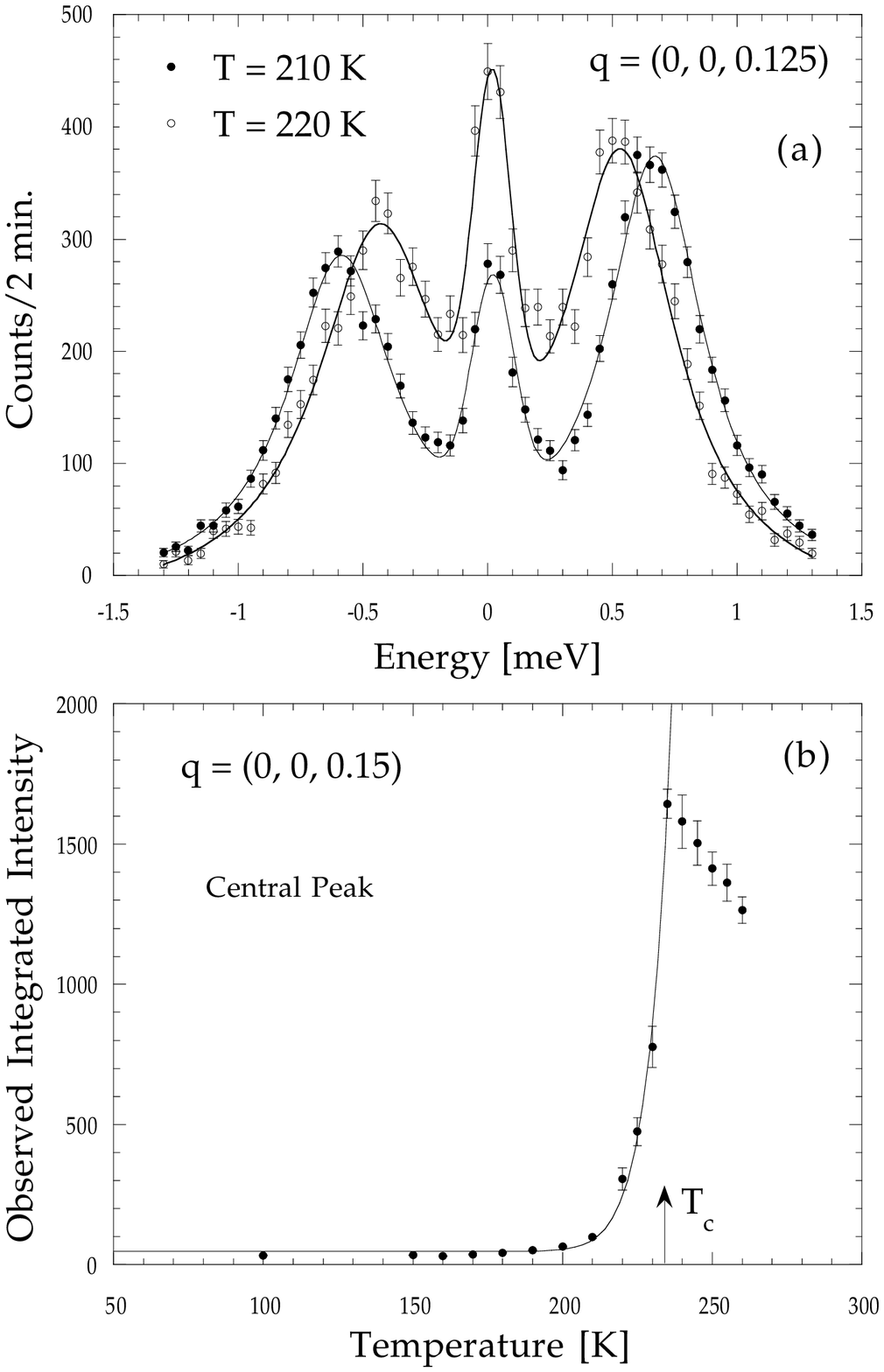,width=17truecm}

\Large
\hspace{8cm} FIG. 6: L. Vasiliu-Doloc et al.

\newpage
\clearpage
\vspace*{0cm}
\hspace*{1.0cm} \epsfig{file=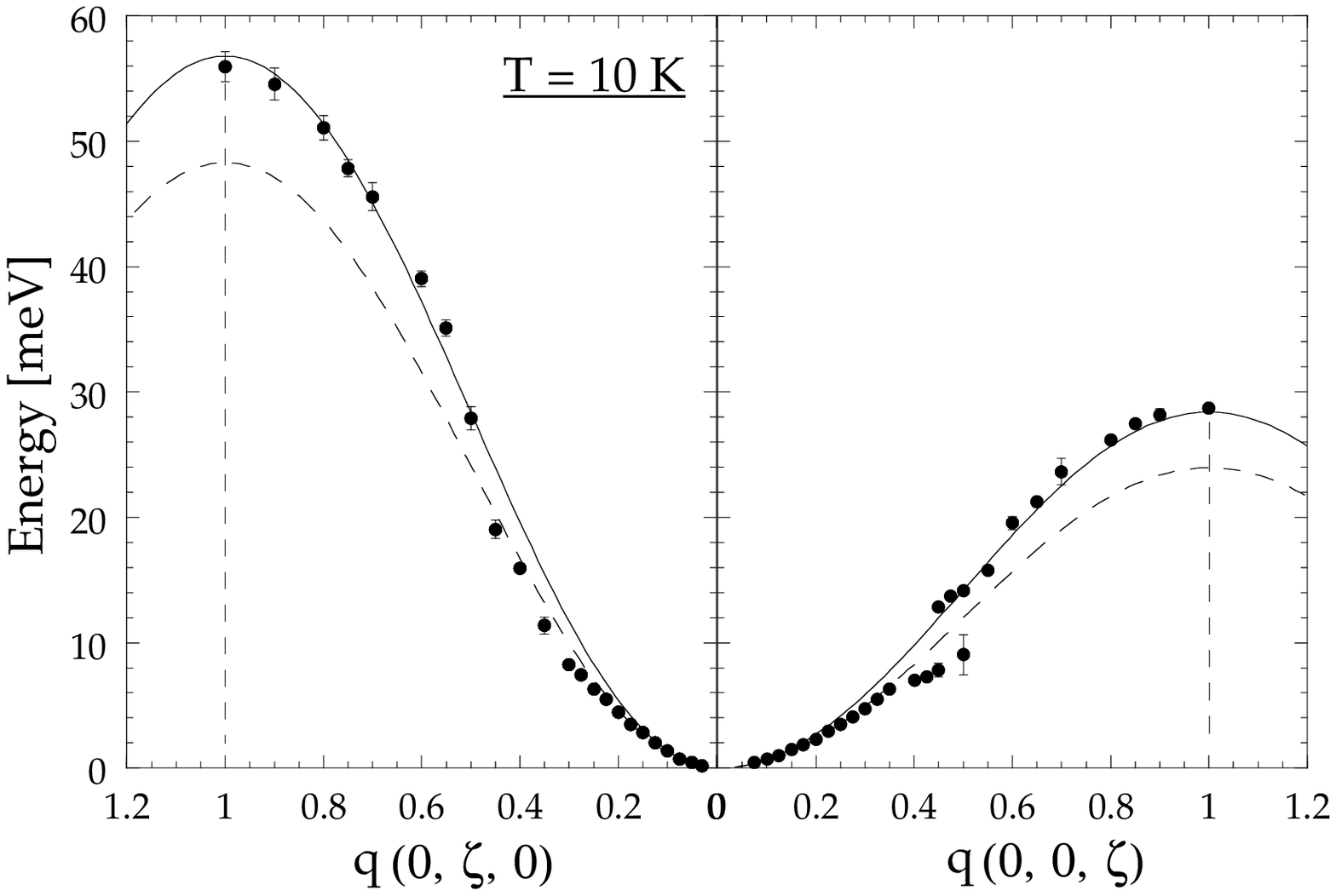,width=17truecm}

\vspace{-2cm}
\Large
\hspace{8cm} FIG. 7: L. Vasiliu-Doloc et al.

\newpage
\clearpage
\vspace*{3.0cm}
\hspace{1cm} \epsfig{file=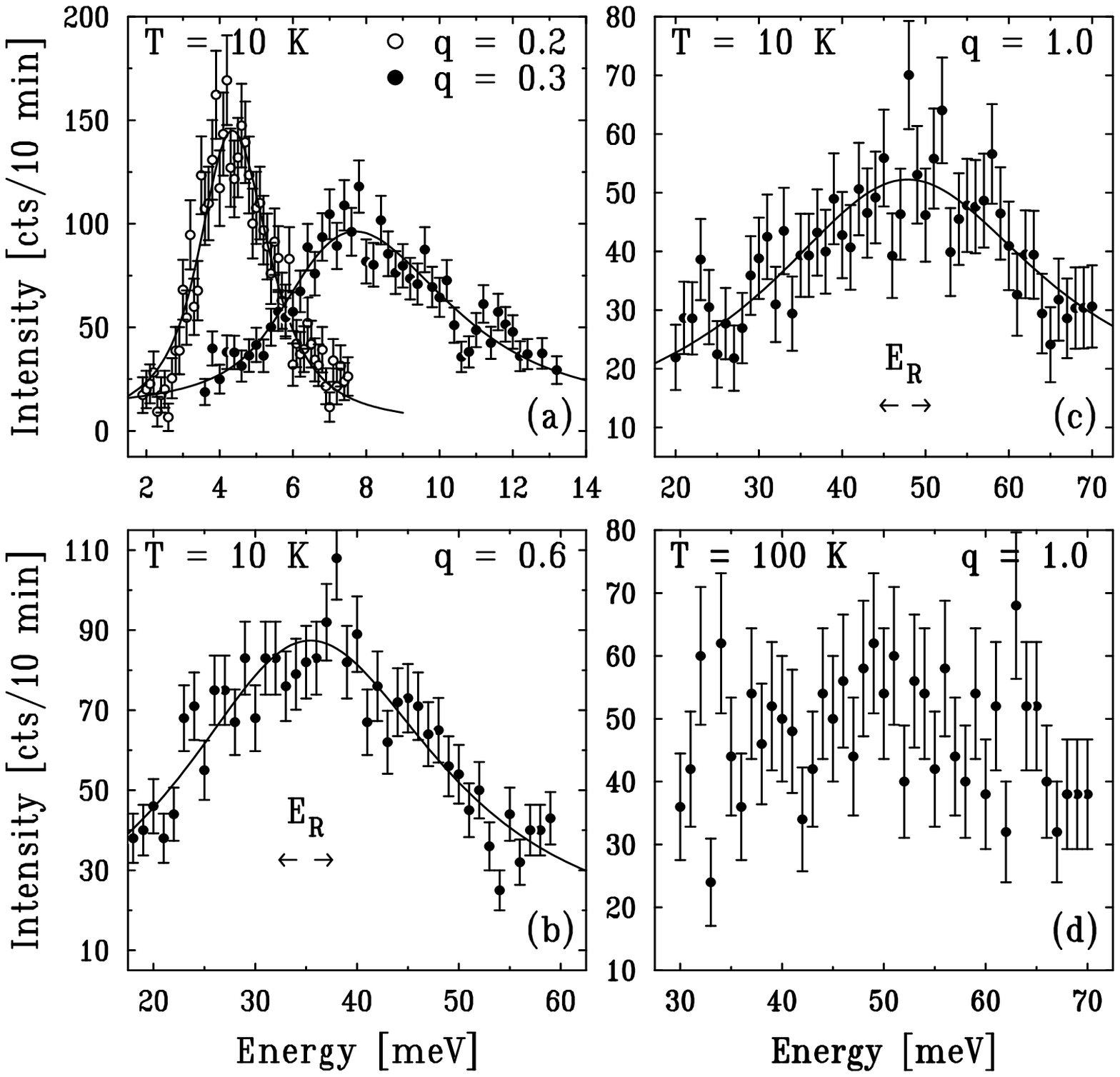}

\Large
\vspace*{2.0cm}
\hspace*{8cm} FIG. 8: L. Vasiliu-Doloc et al.

\newpage
\clearpage
\vspace*{-1.5cm}
\hspace*{1cm} \epsfig{file=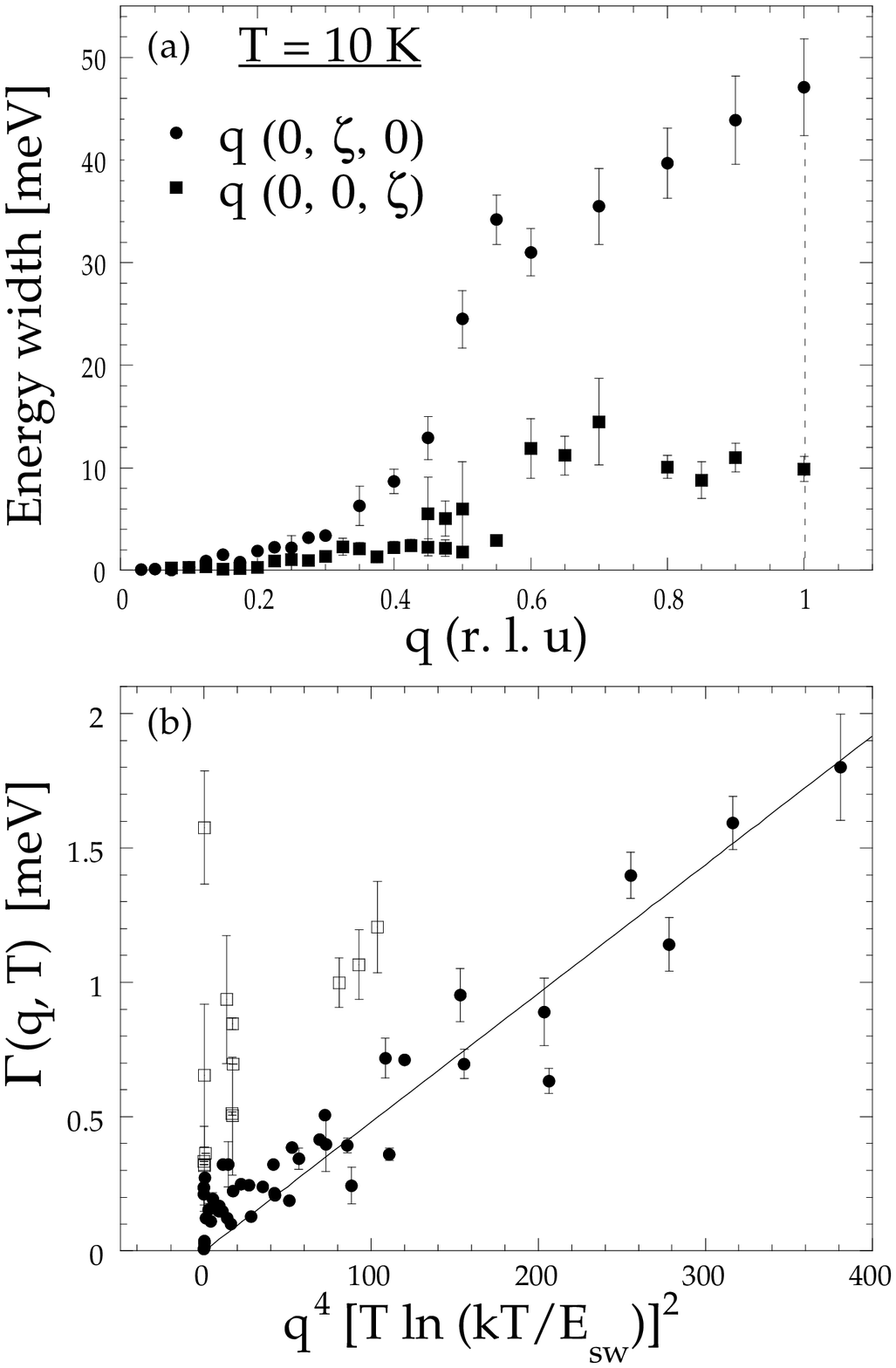,width=17truecm}

\Large
\hspace{8cm} FIG. 9: L. Vasiliu-Doloc et al.


\begin{references}
\bibitem{jonker}  For early literature, see G. H.  Jonker and J. H. van Santen, 
Physica {\bf 16}, 337 (1950); E. O. Wollan and W. C. Koehler, Phys. Rev. 
{\bf 100}, 545 (1955); G. H. Jonker, Physica {\bf 22}, 707 (1956); J. B. 
Goodenough, Phys. Rev. {\bf 100}, 564 (1955).

\bibitem{ecorr}  Y. Tokura et al., J. Phys. Soc. Jpn. {\bf 63}, 3931 (1994).

\bibitem{cmr}  K. Chabara et al., Appl. Phys.  Lett. {\bf 63}, 1990 (1993); S. 
Jin et al., Science {\bf 264}, 413 (1994).

\bibitem{zener}  C. Zener, Phys. Rev. {\bf 82}, 403 (1951); P. W. Anderson
and H. Hasegawa, Phys. Rev. {\bf 100}, 675 (1955); P. G. de Gennes, Phys.
Rev. {\bf 100}, 564 (1955).

\bibitem{millis}  A. J. Millis, P. B. Littlewood, and B. I. Shraiman, Phys.
Rev. Lett. {\bf 74}, 5144 (1995); A. J. Millis, Phys. Rev. B {\bf 55}, 6405
(1997).

\bibitem{martin}  M. C. Martin et al., Phys. Rev. B {\bf 53}, 14 285 (1996).

\bibitem{moudden1}  A. H. Moudden et al., Czech. J. Phys. {\bf 46}, 2163 (1996).

\bibitem{jeff}  J. W. Lynn et al., Phys. Rev. Lett. {\bf 76}, 4046 (1996).

\bibitem{perring}  T. G. Perring et al., Phys. Rev. Lett. {\bf 77}, 711 (1996).

\bibitem{baca} J. A. Fernandez-Baca et al., Phys. Rev. Lett. {\bf 80}, 4012 
(1998).

\bibitem{moudden2}  A. H. Moudden et al., Physica B {\bf 234}-{\bf 236}, 859 
(1997).

\bibitem{hirota}  K. Hirota et al., J. Phys. Soc. Jpn. {\bf 65}, 3736 (1996).

\bibitem{kawano1}  H. Kawano et al., Phys.  Rev. B {\bf 53}, 2202 (1996).

\bibitem{wochner}  P. Wochner, A. H. Moudden, L. Vasiliu-Doloc, J. W.
Lynn, A. Revcolevschi, in preparation.

\bibitem{MMM96} L. Vasiliu-Doloc et al., J. Appl. Phys. {\bf 81}, 5491 (1997).

\bibitem{APS} L. Vasiliu-Doloc et al., Bull. Am. Phys. Soc. {\bf 41}, 529 
(1996); ibid. {\bf 42}, 264 (1997).

\bibitem{uru} A. Urushibara et al., Phys. Rev. B {\bf 51}, 14103 (1995).

\bibitem{anane} A. Anane et al., J. Phys.: Condens. Matter {\bf 7}, 7015 (1995).

\bibitem{kawano2}  H. Kawano et al., Phys.  Rev. B {\bf 53}, R14709 (1996).

\bibitem{yamada} Y. Yamada et al., Phys. Rev. Lett. {\bf 77}, 904 (1996).

\bibitem{dyson}  D. C. Mattis, {\em The theory of magnetism},
Springer-Verlag, Heidelberg, 1981.

\bibitem{jeff2} J. W. Lynn et al., J. Appl. Phys. {\bf 81}, 5488 (1997).

\bibitem{sr30} L. Vasiliu-Doloc et al., J. Appl. Phys. {\bf 83}, 7342 (1998).
press).

\bibitem{pyrochlore} J. W. Lynn, L. Vasiliu-Doloc, M. A. Subramanian, Phys.
Rev. Lett. {\bf 80}, 4582 (1998).

\bibitem{furu1} N. Furukawa, J. Phys. Soc. Jap. {\bf 65}, 1174 (1996).

\bibitem{lor}  We have also used a Lorentzian for the assumed spectral
weight function, and results are qualitatively the same.

\bibitem{Hwang}  H. Hwang et al., Phys. Rev. Lett. {\bf 80}, 1316 (1998).

\bibitem{cooke}  See, for example, J. F. Cooke, J. W. Lynn, and H. L. Davis, 
Phys. Rev. B {\bf 21}, 4118 (1980). 

\bibitem{furu2}  N. Furukawa et al., Physica B {\bf 241}-{\bf 243}, (1998)
(in press).

\bibitem{rush} G. S. Rushbrooke et al., in {\it Phase Transitions and 
Critical Phenomena}, edited by C. Domb and m. S. Green (Academic, New 
York, 1974).

\bibitem{furu3} N. Furukawa, J. Phys. Soc. Jap. {\bf 64}, 2754 (1995).

\end{references}
\end{document}